\documentclass[twocolumn,aps, nofootinbib,showkeys]{revtex4}

\usepackage{amsmath}
\usepackage{graphicx}
\usepackage{dcolumn}
\usepackage{bm}

\begin{document}

\title[]{Peer-review in a world with rational scientists: Toward selection of the average}

\author{Stefan Thurner$^{1,2,3,*}$ and Rudolf Hanel$^1$;}

\affiliation{$^1$Section of Science of Complex Systems, Medical University of Vienna, Spitalgasse 23, A-1090, Austria\\ 
$^2$Santa Fe Institute, 1399 Hyde Park Road, Santa Fe, NM 87501, USA\\
$^3$ IIASA, Schlossplatz 1, A-2361 Laxenburg,  Austria}
\email{stefan.thurner@meduniewien.ac.at}

\begin{abstract}
 One of the virtues of peer review is that it provides a self-regulating selection mechanism for scientific work, papers and projects. 
Peer review as a selection mechanism is hard to evaluate 
in terms of its efficiency. Serious efforts to understand its strengths and weaknesses  have not yet lead to clear answers. 
In theory peer review works if the involved parties (editors and referees) conform to a set of requirements,  such as 
love for high quality science, objectiveness, and absence of biases, nepotism,  friend and clique networks, selfishness, etc. 
If these requirements are violated, what is the effect on the selection of high quality work? We study this question with a 
simple agent based model. In particular we are interested in the effects of {\em rational} referees, who might not have any incentive to 
see high quality work other than their own published or promoted. 
We find that a small fraction of incorrect (selfish or rational) referees can drastically reduce the quality of the published (accepted) scientific standard. 
We quantify the fraction for which peer review will no longer select better than pure chance. 
Decline of quality of accepted scientific work is shown as a function of the fraction  of rational and unqualified referees. 
We show how a simple quality-increasing policy of e.g. a journal can lead to a loss in overall scientific quality, and how mutual support-networks  of authors and 
referees deteriorate the system.



 \end{abstract}

\maketitle

\section{Introduction}
''Dear Sir, We had sent you our manuscript for publication and had not authorized you to show it to specialists before it is printed...'' was  
A. Einstein's reply to the editor of Physical Review  when he received the rejection of a 1936 paper with N. Rosen \cite{pt}. To Einstein peer review  
was not a familiar system at the time. Quality selection of scientific work was dominated by autonomous 
decisions  of editors of unquestionable authority (e.g. Max Plank for {\em Annalen der Physik}) or -- for the  more senior scientists -- by academy memberships. 
Since these days scientific publishing has changed to a practically all dominating peer review system, worldwide and across disciplines. 
Today most journals employ single-blind peer review. 
This transition to  peer review set in after WWII, partly as a response to the increase of scientific research at that time. 

Peer review is a system that subjects scientific work to scrutiny of experts in the field. If the standards of scientific rigor, technical correctness, novelty and the criterion of sufficient interest  
are approved by usually 2-3 peers, the work gets published. In this way the authority of science -- represented by a team of experts of a particular community 
-- self-regulates its  quality standards. Ideally, these standards should be  kept as high as possible. The peer review system is largely perceived as to meet this 
aim, as one of the best possible ways, with -- at present -- little alternatives \cite{naturetrial,doubleblind}.

Peer review suffers from well known problems \cite{groves2006}, maybe the most impressive one being that its efficiency  is largely unknown and practically untested
\cite{jefferson2002}. 
The  Cochrane collaboration has not found evidence that peer review works, nor that it does not \cite{jefferson2007}. 
In an experiment peer review has been tested for its ability to detect errors in scientific papers. 
A  paper with 8 deliberate errors  was sent to 420 reviewers. The average number of detected errors was 2, nobody spotted 5 or more errors, and 16\% did 
not spot a single error \cite{rothwell2000}. Further, peer review leaves room for bias on both the reviewer's 
and editor's side. Evidence for existence of nationality, language, speciality, reputation and gender biases have been reported \cite{godlee1998,wenneras1997}.  

A fundamental problem of the peer review process is that it introduces conflicting interests or moral hazard problems in a variety of situations.
By accepting high quality work  and  thus  promoting it, the referee risks to draw the attention to these ideas and possibly away from her own. 
A post-doc looking for his next position is maybe not happy to accept a good paper of his peer who competes for the same position. 
A big-shot in a particular field might fear to risk his 'guru status'  by accepting challenging and maybe better ideas than his own, etc. 
In other words, referees who optimize their overall 'utility'  (status, papers, fame, position, ...) might find that 
accepting good scientific work of others is in direct conflict with their own utility maximization. 
In the following we call utility optimizing referees {\em rational}. 
They avoid to accept better work than their own.  

It is clear that in the presence of referees with conflicting interests the quality selection aspect of the peer review system will work sub-optimally. 
Here we present a simple agent based model to assess the  robustness of the quality selection component of peer review under the presence of 
rational referees. The choice of an agent based model is motivated by the absence of alternatives to assess efficiency. 
It is possible to demonstrate that the concept of peer review works fine as long as rational referees are absent. However, the system is extremely 
sensitive to perturbations. The model allows to illustrate the impact of rational behavior,  of friendship networks (nepotism), 
and counter-intuitive effects, e.g. from quality-increasing management policies, implemented  by journals or funding agencies.  

Here we solely  focus on the quality selection aspect of peer review and ignore its other potential benefits such as  improvement of papers 
through spell checking, error detection, completing references, cost efficiency, etc. 
Also we will not discuss its inadequacy to select  novel ideas \cite{brezis2007,horrobin1990}, nor its experimentally confirmed 
conformatory bias \cite{mahoney1977, martin1997}. From the quality selection perspective only, we find that unless the fractions of rational and 
unreliable referees are kept well below $30\%$, peer review will not perform better than pure chance, e.g. by throwing a coin. It is interesting to 
see this finding in the light of recently proposed semi-random selection process which has been shown to have a economic {\em value}-adding component 
in the context of R\&D projects \cite{brezis2007}.

\section{Model}

We consider a scientific community of $N$ productive scientists. Each scientist produces one paper every 2 time-units $t$ (e.g. one paper every 2 weeks, years or decades). 
We make two assumptions about the authors: \\

(1) The scientific quality of authors follows a Gaussian distribution, i.e.  
each author $i$ is assigned an 'IQ' index, $Q^{\rm author}_i$, drawn  from a normal distribution $Q^{\rm author}_i \in N(100,\sigma^2_{\rm author})$. 
We consider mature  scientists, i.e.  $Q^{\rm author}_i$ is constant over time (no learning or intellectual decline). We set  $\sigma_{\rm author}=10$.
 
(2) Author $i$ produces papers which vary in quality, fluctuating around his average quality $Q^{\rm author}_i$. We assume these fluctuations to be Gaussian with 
variance $\sigma_{\rm quality}^2$. The quality of a paper submitted by $i$ 
at timestep $t$ is $Q^{\rm submit}_i (t) \in N(Q^{\rm author}_i, \sigma_{\rm quality}^2)$.  For simplicity $\sigma_{\rm quality}=5$ is the same for all authors. 
The distribution of all submitted papers thus is $N(100, \sigma^2_{\rm author}+ \sigma_{\rm quality}^2)$.  
The essence of the review process is now to select the good papers from these submissions. \\

At every timestep each submitted  paper is sent to 2 independent referees, randomly chosen from the $N$ scientists. 
The possibility of self-review is excluded. 
Each reviewer produces a binary recommendation within the same timestep: 'accept' or 'reject'. 
If a paper gets 2 'accept' it is accepted, if it gets 2 'reject', it is rejected, if there is a tie (1 'accept' and 1  'reject') it gets accepted with a probability of $0.5$. 
Upon acceptance a {\em submitted paper} becomes an {\em accepted paper}, the quality of which is now denoted by $Q^{\rm accept}_i (t)$. 
We assume each scientist, when acting as a referee, to belong to one of four referee types: 

\begin{itemize}
\item
{\bf The correct:} Accepts good and rejects bad papers. A paper from author $j$ is considered good by referee $i$, 
if its quality is above a minimum requirement $Q^{\rm min}$. This minimum requirement can be modeled in various ways. Here we
say that the minimum paper quality -- set e.g. by a journal -- is that it lies within the top $q$-quantile of recently published papers in the field. 
To compute this quantile we first use a simple exponential moving average to compute the average: 
$M(t)=\lambda M(t-1) + (1-\lambda)\langle Q^{\rm accept}_i(t-1) \rangle _i $, with $\lambda=0.1$.
$\langle.\rangle_i$ indicates the average over all {\em accepted} papers. The top $q$-quantile is now everything 
above this moving average plus $\alpha$ times one standard deviation of the distribution of recently  published papers, 
$Q^{\rm min}(t)= M(t)+\alpha \, {\rm std}[Q^{\rm accept}_i(t-1)]$.
In other words, referee $i$ accepts a paper if $Q^{\rm submit}_i (t) \geq Q^{\rm min}(t) $.

\item
{\bf The stupid (random):} This referee can not judge the quality of a paper (e.g. because of incompetence or lack of time) 
and  takes a random decision on a paper. 

\item
{\bf The rational:} The rational referee knows that work better than his/her own might draw attention away from his/her own work. 
For him there is no incentive to accept anything better than one's own work, while it might be fine to accept worse quality.   
Referee $i$ rejects papers from authors $j$ whenever they are  above his/her quality index,  $Q^{\rm author}_i < Q^{\rm submit}_j (t)$. 
$i$ accepts papers when they are between a minimum quality threshold and his own index, i.e. 
$Q^{\rm submit}_j (t) \in [Q^{\rm submit}_{\rm min}, Q^{\rm author}_i ]$. 
For simplicity we set $Q^{\rm submit}_{\rm min}=90$ constant.

\item
{\bf The altruist:} Accepts all papers. The referee might simply not care or could fear that his identity might get disclosed through e.g. editorial mistakes. 

\item
{\bf The misanthropist:} Rejects all papers. 

\end{itemize}

\begin{figure}[t]
\begin{center}
	\includegraphics[width=8.0cm]{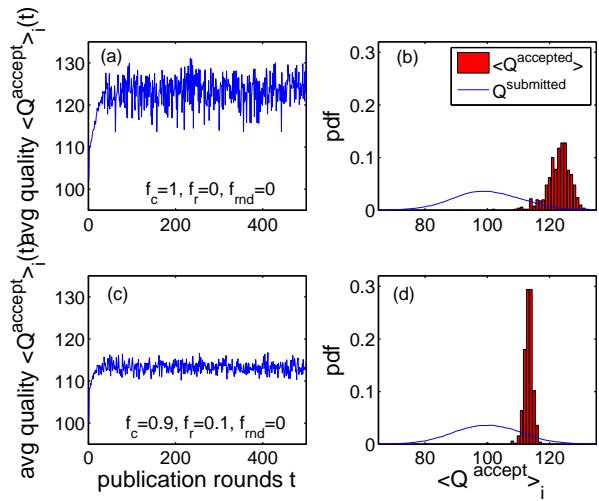}
\end{center}
\caption{Average quality of accepted papers at each timestep for a community of  correct referees (a) and 
one composed  of 10\% rational vs. 90 \% correct referees (c). In (b) and (d)  the histograms of the respective timeseries (a) and (c) are 
shown (red). For reference  we indicate the quality distribution of submitted papers (blue line in (b) and (d)). Note, that the average quality of submitted papers is 100. 
The average paper quality  with 10\% rational  referees drops by about 10, i.e. one  standard deviation of the submitted paper distribution.}
\label{fig1}
\end{figure}

Recommendations of the altruist and misanthropist type affect the total number of accepted papers but do not influence the quality selection process. 
We therefore neglect them and  remain with three types of referees: the correct, the random and the rational, their  
respective fractions being $f_c$, $f_{\rm rnd}$  and $f_{\rm r}$, with $f_c + f_{\rm rnd}+f_{\rm r}  =1$.

To model effects of mutually favoring friendship networks where friends (or members of some group such as co-authors) accept each other's work regardless of 
quality, we introduce a variant  of the model. We randomly chose $L$ scientists who belong to a 
(say co-authorship) network within the community. $N-L$ authors do not belong the network. Regardless of 
referee type, if the author of a submitted paper $i$ and the referee $j$ both belong to the same group, the paper is automatically accepted.  
The relevant parameter is relative group size, $n=L/N$.

The role of journals or funding agencies we model by allowing them to  ask referees 
to accept papers only if they fall within the top $q$-quantile. According to the previous notation,  journals can specify their $\alpha$. 
Unless stated explicitly,  we set $\alpha=0$, i.e. referees are asked to accept above-average papers. 
Note, that only the correct referees will comply with this requirement.

\section{Results}

 We implement  the above model in a computer simulation for $N=1000$ scientists and $500$ timesteps. We first 
set $\alpha=0$, i.e. referees are asked to accept above average articles only.

In Fig. \ref{fig1} we show the time evolution of the average accepted quality of papers at each timestep. In (a) 
the situation is shown for a community of 100\% correct referees. After a short transition time the average paper quality
of accepted papers, rises above 120, i.e. only top quality papers are selected. In  Fig. \ref{fig1} (b) we show the histogram 
of the data points of Fig. \ref{fig1} (a), i.e.  accepted paper quality (red) over 500 timesteps.  It is clear that the average of the 
accepted papers is about 2 standard deviations above the mean of the submitted papers (100). 
The referee process works at its best, only the very best papers are published. 
In Figs. \ref{fig1} (c) and (d) we show the same as above, now for a community composed of 10\% rational and 90\% 
correct referees. It is immediately clear that the average quality of the accepted papers drops drastically --  
the same refereeing process with a small fraction of rational referees (10\%) brings the quality down by almost one standard deviation of submitted article quality. 

 \begin{figure}[t]
\begin{center}
	\includegraphics[width =7.0cm]{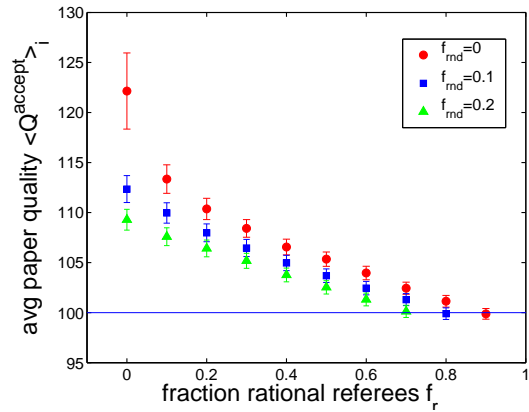}
\end{center}
\caption{Average quality of accepted papers as a function of the abundance of  rational referees. The three different scenarios shown correspond to three
fractions  of random referees, $f_{rnd}=0,0.1$ and $0.2$. It is obvious that even small fractions of rational referees bring down the system to select 
papers of close to average quality. $\alpha=n=0$.
}
\label{fig2}
\end{figure}

The decline of average accepted paper quality as a function of  the fraction of rational referees in the community is shown in Fig. \ref{fig2}. 
For fractions of over 70\% of rational referees (an admittedly dark scenario) the selection mechanism in the refereeing process completely vanishes. 
The average quality  of selected work falls within a narrow band  of average quality (100) and peer review process turnes absurd. 
Fig. \ref{fig2} further shows the situation for two fixed fractions of random referees, $f_{rnd}=0.1$ and $0.2$. Clearly, the addition of random referees brings the 
accepted paper quality down in a dramatic way. Quantitatively, adding about 10\% of random referees has practically the same effect as adding 10\% of rational referees, 
see Fig. \ref{fig2}.

\begin{figure}[t]
\begin{center}
	\includegraphics[width=7.0cm]{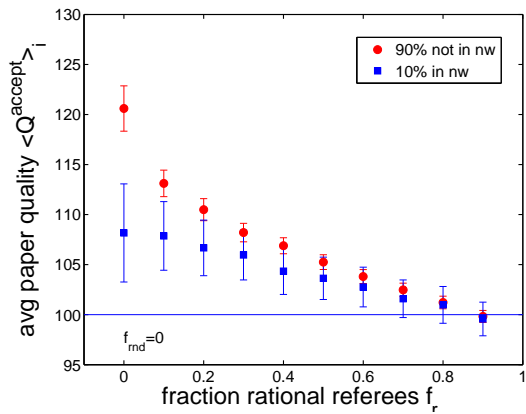}
\end{center}
\caption{Situation with a subgroup of authors ($n=0.1$) who accept all submissions from authors belonging to the 
same group.  The average accepted paper quality of the authors from the group (squares)  is compared to the one of those 
 not belonging to the cartel (circles), again shown as a function of rational referees, $f_r$. 
Obviously, when the refereeing system is surpassed by a 'friendship-bias' no quality selection is possible. $f_{rnd}=0$, $\alpha=0$. }
\label{fig4}
\end{figure}

\begin{figure}[t]
\begin{center}
	\includegraphics[width=7.0cm]{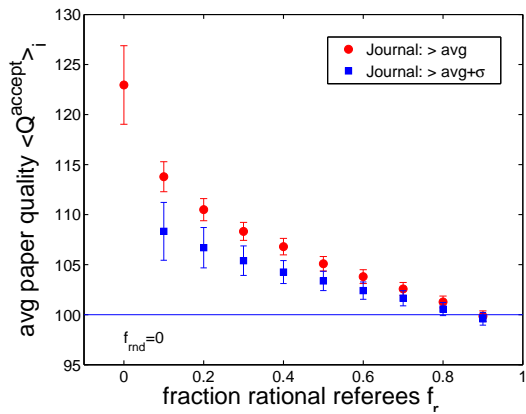}
\end{center}
\caption{Effect of journals asking for higher paper quality standards.  The cases $\alpha=0$ (circles) and  $\alpha=1$ (squares) are compared, which 
is a shift of one standard deviation. 
The later leads to a significant drop of average accepted paper quality.  $f_{rnd}=n=0$.}
\label{fig3}
\end{figure}

In  Fig. \ref{fig4} we show the results from a simulation with a friendship network present, whose members 
accept all papers of authors belonging to the same network, regardless of paper quality or referee type. 
We set $n=0.1$, i.e. 10\% of the authors are part of such a network. 
Again we plot the scenario as a function of abundance of rational referees, $f_{rnd}=0$, and $\alpha=0$.
The figure compares the average quality of papers of authors belonging to the network (squares), 
to those of non-members (circles). As expected, the quality of members of the network is drastically lower than for the rest where 
the referee system is not corrupted. The average quality of accepted papers in the total society  
drops from about $122$ to $117$ for $f_{r}=0$. 
The acceptance rate of authors belonging to the network is about a factor of 2  larger than for the correct group, irrespective of the fraction 
of rational referees. In absolute terms the acceptance rate for the authors outside the network is 
about 5\%. In the survey \cite{ware2008} a rate of acceptance (within the first round) of about 8\% was reported, placing the model results 
in the right ballpark.

Finally, in Fig. \ref{fig3} we show the effect of the journal asking referees to raise the quality threshold, i.e. 
$\alpha>0$. The circles reflect the situation as in  Fig. \ref{fig2} (for $f_{rnd}=0$).  
Once referees are asked to raise their threshold from $\alpha=0$  by one standard deviation to $\alpha=1$, quality drops (squares). 
The reason is easily understanable: while correct referees, by raising their standards, accept less and less (good) papers, 
the bias toward bad papers of the rational referees remains the same. 
The rational referees gain more 'weight' in the selection process with respect  to the correct ones, and the quality deteriorates.

\section{Conclusion}

With a simple agent based model we have shown that the standard peer reviewing process is not a robust way for quality selection of scientific work. 
The presence of relatively small fractions of  'rational' and/or 'random' referees (deviating from correct behavior) considerably reduces the average quality of 
published or sponsored science as a whole.
We quantified the effect of nepotism networks and demonstrated that under the presence of 'rational'  
referees quality-increasing strategies of journals or funding agencies can lead to adverse 
effects on a systemic level. 
Our message is clear: if it can not be guaranteed  that the  fraction of 'rational'  and 'random' referees is confined to a very small number, the peer review
system will not perform much better than by accepting papers by throwing (an unbiased!) coin. 
For example  if the fractions of rational, random and correct referees are approximately 1/3 each, the quality selection aspect of peer review practically vanishes. 
We think that it is important to try to think of ways to assess the actual numbers within the different scientific communities. 
If it turns out that e.g. $f_{\rm rnd}$ and $f_{r}$  fall above 30\% it would be necessary to re-think the prevailing system. 
Under these circumstances -- which are not totally unrealistic for certain communities --  a purely random refereeing system would perform  
equally well --  and would at the same time  safe millions of man-hours spent on refereeing every year.

We thank Elise S. Brezis for most helpful comments.


\begin{thebibliography}{99}

\bibitem{pt}
Kennefick, D. 
Einstein Versus the Physical Review. 
Physics Today {\bf 58} 43 (2009). 

\bibitem{naturetrial}
Nature's peer review trial, 
Nature 05535 (2006),\\
www.nature.com/nature/peerreview/debate/nature05535.html; 
Editorial: Peer review and fraud, 
Nature {\bf 444}, 971-972 (2006).

\bibitem{doubleblind}
Editorial:  Working double-blind: should there be author anonymity in peer review? 
Nature {\bf 451}, 605-606 (2008).

\bibitem{groves2006}
Grove, T.
Quality and value: how can we get the best out of peer review?
Nature (Nature Peer Review debate) 04995 (2006), \\
www.nature.com/nature/peerreview/debate/nature04995.html 

\bibitem{jefferson2002}
Jefferson, T.,  Alderson, P., Wagner, E., Davidoff, F.  
Effects of editorial peer review: A systematic review.
J. of the Am. Med. Assoc. {\bf 287},  2784-2786 (2002).

\bibitem{jefferson2007}
Jefferson, T., Rudin, M., Brodney Folse, S., Davidoff, F.
Editorial peer review for improving the quality of reports of biomedical studies. 
Cochrane database of systematic reviews 2007 {\bf 2},  MR000016 (2007).

\bibitem{rothwell2000}
Rothwell, P.M.,  Martyn, C.N.  
Reproducibility of peer review in clinical neuroscience: is agreement between reviewers any greater than would be expected by chance alone? 
Brain {\bf 123}, 1964-1969 (2000).

\bibitem{godlee1998} 
Godlee, F., Gale, C.R., Martyn, C.N.  
The effect on the quality of peer review of blinding reviewers and asking them to sign their reports: a randomised controlled trial.
J. of the Am. Med. Assoc. {\bf 280}, 237-240 (1998).

\bibitem{wenneras1997}
Wenneras, C., Wold A.  
Nepotism and sexism in peer review. 
Nature {\bf 387}, 341Ð343 (1997).

\bibitem{brezis2007}
E.S. Brezis, 
Focal randomization: an optional mechanism for the evaluation of R\&D projects. 
Science and Public Policy {\bf 34},  691-698 (2007).

\bibitem{horrobin1990}
Horrobin, D.F.  
The philosophical basis of peer review and the suppression of innovation. 
J. of the Am.  Med. Assoc. {\bf 263}, 1438Ð1441 (1990).

\bibitem{mahoney1977}
Mahoney, M.  
Publication prejudices: an experimental study of confirmatory bias in the peer review system. 
Cognitive Therapy and Research {\bf 1}, 161Ð175 (1977).

\bibitem{martin1997}
Martin, B.  
Peer review as scholarly conformity. 
In Suppression Stories,  pp. 69-83. Wollongong: Fund for Intellectual Dissent (1997).

\bibitem{ware2008}
Ware, M. 
Peer review: benefits, perceptions and alternatives.   
Publishing Research Consortium Summary Papers {\bf 4}, 1-20 (2008).


\end{thebibliography}
\end{document}